\documentclass[runningheads]{llncs}
\usepackage[T1]{fontenc}
\usepackage{graphicx}
\usepackage{subcaption}
\usepackage{caption}
\usepackage{amsmath}
\usepackage{amssymb}
\begin{document}
\title{AgentZero++: Modeling Fear-Based Behavior }
\author{Vrinda Malhotra \and
Jiaman Li\ \and
Nandini Pisupati}
\authorrunning{V. Malhotra et al.}
\institute{George Mason University, Fairfax VA 22030, USA 
\\email{vmalhot2@gmu.edu; jli30@gmu.edu; npisupat@gmu.edu}}
\maketitle              % typeset the header of the contribution
\begin{abstract}
We present AgentZero++, an agent-based model that integrates cognitive, emotional, and social mechanisms to simulate decentralized collective violence in spatially distributed systems. Building on Epstein’s Agent\_Zero framework, we extend the original model with eight behavioral enhancements: age-based impulse control, memory-based risk estimation, affect-cognition coupling, endogenous destructive radius, fight-or-flight dynamics, affective homophily, retaliatory damage, and multi-agent coordination. These additions allow agents to adapt based on internal states, previous experiences, and social feedback, producing emergent dynamics such as protest asymmetries, escalation cycles, and localized retaliation. Implemented in Python using the Mesa ABM framework, AgentZero++ enables modular experimentation and visualization of how micro-level cognitive heterogeneity shapes macro-level conflict patterns. Our results highlight how small variations in memory, reactivity, and affective alignment can amplify or dampen unrest through feedback loops. By explicitly modeling emotional thresholds, identity-driven behavior, and adaptive networks, this work contributes a flexible and extensible platform for analyzing affective contagion and psychologically grounded collective action.

\end{abstract}
\section{Introduction}
Nietzsche famously remarked that “madness is the exception in individuals but the rule in groups,” capturing a philosophical intuition that has since shaped the foundational theories of psychology and social thought \cite{Nietzsche2002}. As psychology emerged as a scientific discipline in the late 19th century, scholars began systematically investigating how individuals behave differently in collective settings. Gustave Le Bon, writing shortly after Nietzsche, gave an empirical form to this idea by arguing that factors such as anonymity and suggestibility amplify contagion within crowds, potentially causing them to behave as if they possess their 'own minds' \cite{LeBon1895}. Neuroscience research suggests that fear responses are deeply embedded in brain architecture, particularly involving structures such as the amygdala and prefrontal cortex \cite{Bloom2006}.

The concept of contagion reappears prominently in Elias Canetti’s Crowds and Power (1984), where he explores how the pursuit of power or the instinct for survival can drive individuals in crowds to behave irrationally \cite{Canetti1984}. He argues that crowds thrive on density, possess an inherent drive to grow, and tend to move with singular purpose and direction. Canetti’s focus on the psychological structure of crowds is echoed in Christopher Browning’s Ordinary Men (1998), which further emphasizes the power of social norms and conformity within group dynamics, especially in authoritarian contexts \cite{Browning1998}. Through a case study of a German Reserve Police Battalion operating under Nazi rule, Browning argues that many of the men who carried out atrocities were not driven by ideological zeal, but rather by a susceptibility to peer pressure and a willingness to defer to authority. Their moral judgments, he suggests, were reshaped to align with the prevailing norms of the group.

These ideas are further reinforced by findings in social psychology, most notably in Stanley Milgram’s (1974) obedience experiments and Philip Zimbardo’s (2007) Stanford Prison Experiment, both of which demonstrated that ordinary people can adopt roles and behaviors that conflict with their personal ethics when subjected to group pressure or institutional authority \cite{Milgram1974,Zimbardo2007}. Albert Bandura’s work on moral disengagement also sheds light on how individuals rationalize harmful behavior by diffusing responsibility or dehumanizing targets \cite{Bandura1999}. These empirical findings helped lay the foundation for modern theories of collective behavior that incorporate not only emotional contagion but also the cognitive processes behind moral collapse and social influence.

It is because of cases like this that understanding collective unrest's cognitive and social roots is both a key part of our understanding of society and why it remains a core challenge in generative social science, especially when it emerges spontaneously or irrationally. Joshua Epstein's \textit{Agent\_Zero: Toward Neurocognitive Foundations for Generative Social Science} uses the aforementioned psychological theories and many more to introduce a cognitively plausible agent framework composed of three interdependent modules: affective learning, deliberative reasoning, and social influence through dispositional contagion. One of his key illustrations, \textit{Parable 1: The Slaughter of Innocents Through Dispositional Contagion}, demonstrates how agents may act destructively when their internal disposition surpasses a threshold without copying others' behavior or following orders\cite{Epstein2013}.

We construct a hybrid model that incorporates the cognitive architecture of Agent\_Zero (as shown in Parable 1) within the structural setting of the civil violence model. This allows us to simulate not only how fear and belief drive behavior but also how mobility, spatial distribution, and enforcement mechanisms interact with cognition. Our study aims to evaluate how adding cognitively grounded extensions to this hybrid model alters the dynamics of fear-based collective behavior. Our central research question is: How do affective, cognitive, and social mechanisms—augmented with memory, age, behavioral heuristics, and identity bias—change group-level outcomes in a mobile protest environment?

To answer this question, we implemented a Python version of Parable 1 and extended it using eight of Epstein's proposed model enhancements. These extensions help agents behave in ways that are closer to how real people act. For example, agents can now differ by age, change their behavior depending on their fear, choose to run away in some situations, and respond based on past negative experiences. Our simulation results show that these changes strongly affect the overall system. The outbreaks of violence happen at different times, spread across different areas, and often follow a repeating pattern of revenge. In this way, we improve Agent\_Zero and make it a better tool for studying how group violence can grow from individual emotions and decisions.

\section{Literature Review}
Before developing the Agent\_Zero model, Joshua Epstein's earlier works laid a foundational framework for agent-based modeling (ABM) as a method for social science inquiry. In Agent-Based Computational Models and Generative Social Science \cite{Epstein1999}, Epstein introduced the concept of "generative sufficiency" arguing that to explain a social phenomenon truly, one must be able to "grow" it in a computational simulation using rule-based agents. This methodological view has profoundly shaped ABM research within the computational social sciences. In \textit{Nonlinear Dynamics, Mathematical Biology, and Social Science} \cite{Epstein1997}, Epstein further explored dynamical systems and feedback mechanisms to understand population-level behavior, which inspired subsequent models involving contagious social processes.

Agent\_Zero (2013) marked a significant advance by incorporating neurocognitive realism into ABM. Epstein formalized a modular agent structure based on three interacting systems: adaptive learning, deliberative reasoning, and social contagion \cite{Epstein2013}. This triadic model allowed ABMs to capture human behavior more plausibly. A notable innovation was the concept of dispositional contagion, in which agents are influenced not only by others' actions but also by their perceived emotional states. This mechanism allowed the model to generate complex emergent behaviors, such as spontaneous collective violence, without relying on leadership structures or normative conformity.

Epstein uses parables to demonstrate model dynamics consistently in his work. Parable 1 in Agent\_Zero exemplifies this strategy, illustrating how simple cognitive-emotional rules can generate macro-level social patterns \cite{Epstein2013}. The complexity of the model emerges from localized interactions among cognitively heterogeneous agents. A similar logic underlies Epstein's earlier Civil Violence Model \cite{Epstein2002}, in which agents rebel based on a calculus of grievance and state repression. Although different in cognitive sophistication and trigger mechanisms, both models emphasize the importance of internal thresholds and decentralized interactions.

To explore the implications of Epstein's work more deeply, our study incorporates and extends eight of the fourteen cognitive-behavioral modifications Epstein proposed for Agent\_Zero. These include: (1) Age-based Impulse Control, (2) Endogenous Destructive Radii, (3) Fight-or-Flight Dynamics, (4) Introduction of Memory, (5) Retaliatory Damage, (6) Coupling of Affect, (7) Affective Homophily and Endogenous Network Dynamics, and (8) Expansion to Multiple Agents.

For example, Epstein draws on literature in developmental psychology to link youth with lower impulse control, proposing that age can be modeled through a gap between emotional activation and destructive behavior \cite{Mischel2011}. The endogenous destructive radius adds realism by scaling an agent's action radius based on emotional intensity, reducing parameter dependence, and reflecting observed behavior in collective events like vaccine refusal or riot contagion.

The fight-or-flight mechanism, rooted in biological and psychological research on fear responses \cite{deGelder2004}\cite{Baranski2010}, introduces a richer behavioral spectrum, allowing agents to flee rather than fight when fear is high but perceived threat remains manageable. We adapted this into our Python-based model to explore the effects of avoidance behavior on overall conflict intensity. We also isolated the concept of retaliatory damage, briefly mentioned by Epstein, to model feedback loops in aggression inspired by research on excitation and violence \cite{Zillmann1975}.

Episodic memory is another extension that Epstein introduced to simulate agents who learn from immediate and past experiences \cite{bouton2004context}. This is crucial for modeling how perception and disposition evolve. Similarly, Epstein's coupling of affect and cognition reflects dual process theories in psychology \cite{cacioppo1982need}, positing that emotion and rational assessment interact continuously to shape behavior. His use of mathematical bias functions to model this interaction underscores the non-linearity of behavioral outcomes.

Affective homophily, the tendency of agents to associate with emotionally similar others, is operationalized through dynamical updating of social bonds. This mechanism is grounded in network science \cite{barabasi1999emergence} and social neuroscience \cite{cacioppo2007social}, and it enables the model to simulate emergent clustering and polarization. While Epstein did not elaborate extensively on the expansion to multiple agents, we include this in our implementation to explore the collective effects of cognitive heterogeneity in larger populations.

Beyond Agent\_Zero, we also draw from Epstein's Civil Violence model to structure the spatial dynamics of protest and enforcement \cite{Epstein2002}. In this model, the risk of arrest and the density of rebellion co-evolve to produce cycles of unrest. By hybridizing this spatial logic with Agent\_Zero's cognitive architecture, we provide a more comprehensive framework for studying emotionally driven, decentralized violence.

Recent ABM research has expanded on these ideas by incorporating affective dimensions into polarization, radicalization, and rumor-spread models. For example, Flache et al. (2017) highlights the limitations of purely rational models and argue for integrating cognitive-affective dynamics \cite{flache2017models}. Our study contributes to this growing literature by demonstrating how cognitive extensions interact and combine to alter macro-level outcomes.

Rather than exploring a broad range of literature, our work follows a focused, iterative expansion of Epstein's research program. We precisely test how selected extensions influence simulated behaviors and investigate the cognitive and emotional processes that give rise to complex social phenomena.

\section{Methodology}
This simulation combines cognitive-affective dynamics with spatial and social contagion mechanisms to explore the emergence of destructive behavior in a dynamic environment. The model draws from two foundational agent-based models: the spatial mechanics of Epstein’s Civil Violence Model and the cognitive-emotional structure of \textit{Parable 1: The Slaughter of Innocents through Dispositional Contagion}. While the original Parable 1 is implemented in NetLogo, our model is rebuilt and extended using the Mesa framework in Python 3.11 (Mesa version 1.1.0), enabling modular cognitive-behavioral extensions and more sophisticated data handling.

The model operates on a 50×50 toroidal grid populated by agents of a generalized AgentZero type. Each agent perceives and acts on its environment based on a three-part internal architecture: affective learning, probabilistic risk estimation, and social contagion. We begin with a minimal version that contains three agents, one immobile and two mobile, as in the original parable. We later expand this to simulations with up to 20 agents to examine scaling dynamics and emergent social clustering. Simulations typically run for 1000 steps, with fixed random seeds ensuring reproducibility.

\subsection{Spatial Environment}
The grid is composed of discrete patches, each representing a unit of space that can exist in one of three states: yellow, indicating an inactive or benign condition where no immediate threat is present; orange, signaling that the patch is in an active and potentially threatening state, possibly due to hostile agents or escalating tension; and dark red, denoting that the patch has been attacked or destroyed, reflecting a state of severe disruption or damage. These states allow the model to capture dynamic shifts in the environment over time.

Mobile agents perform random walks through this environment, observing the activation state of surrounding patches within a defined spatial sampling radius. Based on a configurable attack rate, a subset of yellow patches becomes orange at each tick, simulating the stochastic appearance of local threats. Orange patches revert to yellow based on an extinction rate, enabling de-escalation cycles.
\begin{figure}
\includegraphics[width=\textwidth]{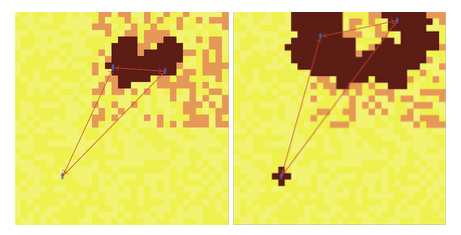}
\caption{Netlogo Model: Activation by Dispositional Contagion.} \label{fig1}
\end{figure}
\subsection{Internal Agent Architecture}
Agents maintain three key internal states: \textbf{Affect} ($A_i$), \textbf{Probability} ($P_i$), and \textbf{Disposition} ($D_i$). 
Affect is updated using a Rescorla-Wagner learning rule, capturing emotional responses to recent exposure to threatening (orange) patches. 
Probability represents the agent’s perceived threat, calculated as a rolling average over a defined memory window, enabling the detection of temporal patterns. 
Disposition combines affect, probability, and socially transmitted influence from connected peers. 
If an agent’s disposition exceeds a behavioral threshold—modulated by its impulse control—the agent takes destructive action.

This framework is formalized as:

\[
D_i = A_i + P_i + C_i - \theta_i(\gamma)
\]

where $C_i$ represents the influence exerted by socially connected peers through network-based contagion, capturing how an individual’s behavior may be shaped by the actions and attitudes of those they are connected to; and $\theta_i(\gamma)$ denotes an individual-specific threshold that depends on the parameter $\gamma_i$ which encodes age-based impulse control. This threshold reflects psychological differences in how readily individuals respond to external stimuli, with variations in $\gamma$ accounting for developmental or experiential factors that affect susceptibility to influence and reactivity to social pressure.

\subsection{Simulation Loop and Logic}
Each simulation tick begins with an environmental update: yellow patches are probabilistically activated to orange based on the attack rate, creating localized threats. Mobile agents observe these threats within their sampling radius and update their internal affect accordingly. Probability estimates are then recalculated using a memory-weighted average of recent environmental conditions, introducing temporal depth to threat perception.

Social influence is transmitted through a directed network of agent connections, where the strength of each tie depends on affective similarity. These weighted links determine how much each agent’s disposition is influenced by its peers. The agent’s final disposition, as a composite of affect, probability, and contagion, is compared to a threshold determined by its impulse control. If the disposition exceeds the threshold, the agent engages in a destructive action, turning patches within its action radius to a destroyed (dark red) state. Retaliatory damage is then incurred based on the reactivity of harmed neighbors.

Throughout the simulation, key agent-level variables—affect, probability, and disposition—are plotted as time-series data to track emergent behavioral dynamics and decision-making patterns.

\subsection{Behavioral Extensions}
\subsubsection{Age and Impulse Control}
Each agent is assigned an age drawn from a Gaussian distribution. Impulse control is computed as:
  \[
  \gamma = 1.0 - \frac{(\text{age} - 18)}{100}
  \]
This inversely correlates age with impulsivity, reflecting psychological evidence that younger people are more susceptible to reactive behavior. Impulse control modulates key behavioral thresholds and amplifies heterogeneity in action. 

\subsubsection{Endogenous Destructive Radius}
In the base Agent Zero model, agents respond to perceived grievance or threat by activating. We extend this by introducing a destructive radius that scales with affect. Formally, the radius $r_d$ is defined as:
  \[
  r_d = \left\lfloor 1 + 4 \cdot \textit{affect} \right\rfloor
  \]
This modeled the idea that emotionally heightened agents have a larger sphere of destructive influence, aligning with real-world observations of collective contagion in unrest and riot dynamics.
\subsubsection{Flight vs. Fight Dynamics}
In addition to the ``FIGHT'' and ``QUIET'' modes, agents with high perceived risk but low disposition can enter a ``FLIGHT'' mode, relocating to safer grid regions. This models avoidance behaviors and adds a third behavioral state to the response spectrum.
\subsubsection{Temporal Memory}
Agents maintain a rolling memory vector of perceived neighborhood activation, introducing \textbf{temporal depth} to their cognitive perception. At each step, agents update their estimate of risk using:
  \[
  \textit{probability}_t = \frac{1}{m} \sum_{i=1}^{m} \textit{active\_neighbors}_{t - i}
  \]
where $m$ is the memory length. This enables agents to base their decisions not only on instantaneous perception but also on \textbf{patterned experience}, aligning the model closer to human behavioral tendencies such as trend learning and desensitization.
\subsubsection{Retaliatory Damage}
Retaliatory damage models the \textbf{feedback cost} an agent incurs after harming others. It reflects the idea that aggression can provoke negative consequences, such as stress, backlash, or retribution.
When agent $i$ attacks neighbors, it inflicts damage based on its \textit{impulse control}: $D_{i \rightarrow j}(t) = \alpha \cdot (1 - IC_i)$, where $\alpha$ is an aggression scaling constant and $IC_i \in [0, 1]$ denotes agent impulse control $i$, with lower values indicating higher impulsivity and thus greater damage inflicted.

In return, agent $i$ accumulates retaliatory damage from each neighbor $j$ it harms, given by $R_i(t) = \sum_{j \in N_i} \mathbb{1}_{\text{harmed}} \cdot \beta \cdot (1 - IC_j)$, where $\beta$ is the retaliation sensitivity constant; $\mathbb{1}_{\text{harmed}}$ is an indicator function (1 if agent $j$ was harmed at time $t$), and retaliatory damage increases if neighbors are highly reactive ($IC_j$ low). This mechanism introduces behavioral \textbf{realism} and a \textbf{self-regulating dynamic} that discourages unchecked escalation.

\subsubsection{Coupling of Affect and Cognition}
Dispositions $D$ are calculated as a function of affect $A_i$, perceived risk $P_i$, and network-based contagion $C_i$, minus a threshold $\theta$ modulated by impulse control:
  \[
  D_i = A_i + P_i + C_i - \theta_i(\gamma)
  \]
By explicitly linking emotion and belief formation in the disposition equation, the model supports tipping-point dynamics and mutually reinforcing feedback between perception and arousal. This supports theories of motivational salience, where the amygdala integrates emotional relevance from personal goals and traits \cite{cunningham2012motivational}.

\subsubsection{Affective Homophily and Endogenous Network Dynamics}
We endogenize the \textbf{strength of social ties} between agents based on affective similarity. At each timestep, link weights $w_{ij}$ in the network are updated as:
  \[
  w_{ij}^{(t+1)} = w_{ij}^t + \alpha \cdot (1 - |A_i - A_j|)
  \]
where $\alpha$ is a learning rate. This models \textbf{affective homophily}, where agents reinforce ties to emotionally similar peers and weaken others. This mechanism reflects real-world clustering by emotional state and generates dynamic networks that evolve with simulation history.
\subsubsection{Expansion to Multiple Agents}
We generalize the model beyond the original 3-agent structure, running simulations with larger populations to explore scale effects and emergent group-level behaviors such as polarization, clustering, and diffusion.

\begin{figure}
\includegraphics[width=\textwidth]{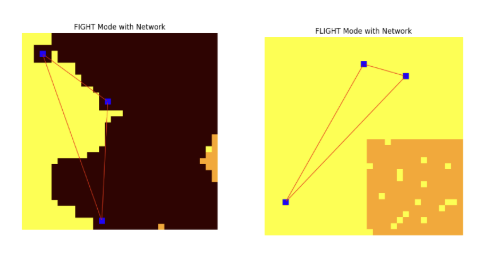}
\caption{Sample of fight vs. flight dynamics shows the difference in coverage.} \label{fig2}
\end{figure}

\begin{figure}
\centering
\includegraphics[width=0.7\textwidth]{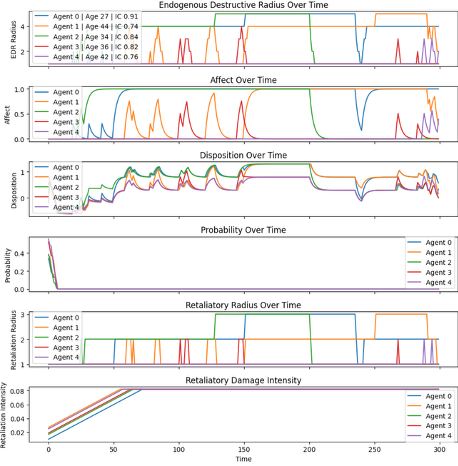}
\caption{Our basic model with five agents and the following additions: age and impulse control, endogenous destructive radius, and retaliatory damage.} \label{fig3}
\end{figure}

\section{Results}

To evaluate the dynamics of cognitive-affective escalation and social contagion, we conducted a series of simulation experiments using the Mesa-based \textit{AgentZero++} model. We implemented and visualized the effects of six key behavioral extensions: \textit{age-based impulse control}, \textit{endogenous destructive radius}, \textit{fight-or-flight behavior}, \textit{retaliatory damage}, \textit{temporal memory}, and \textit{affect-contagion coupling}. Results are presented through spatial snapshots, time-series plots, and network metrics, allowing us to trace how micro-level variation scales into emergent group behavior.

In the base configuration with three agents, destruction occurs when disposition exceeds a fixed threshold driven by affect, probabilistic risk, and peer contagion. Agents 1 and 2, being mobile, respond to environmental threats, while the stationary Agent 0 remains passive. This results in \textit{localized waves of destruction}, with short-lived spikes in affect followed by decay—consistent with emotional extinction learning. Figure~\ref{fig2} illustrates how destructive episodes are spatially contained and temporally brief in the baseline setup.

\begin{figure}[htbp]
  \centering
  \begin{subfigure}[b]{0.7\textwidth}
    \includegraphics[width=\textwidth]{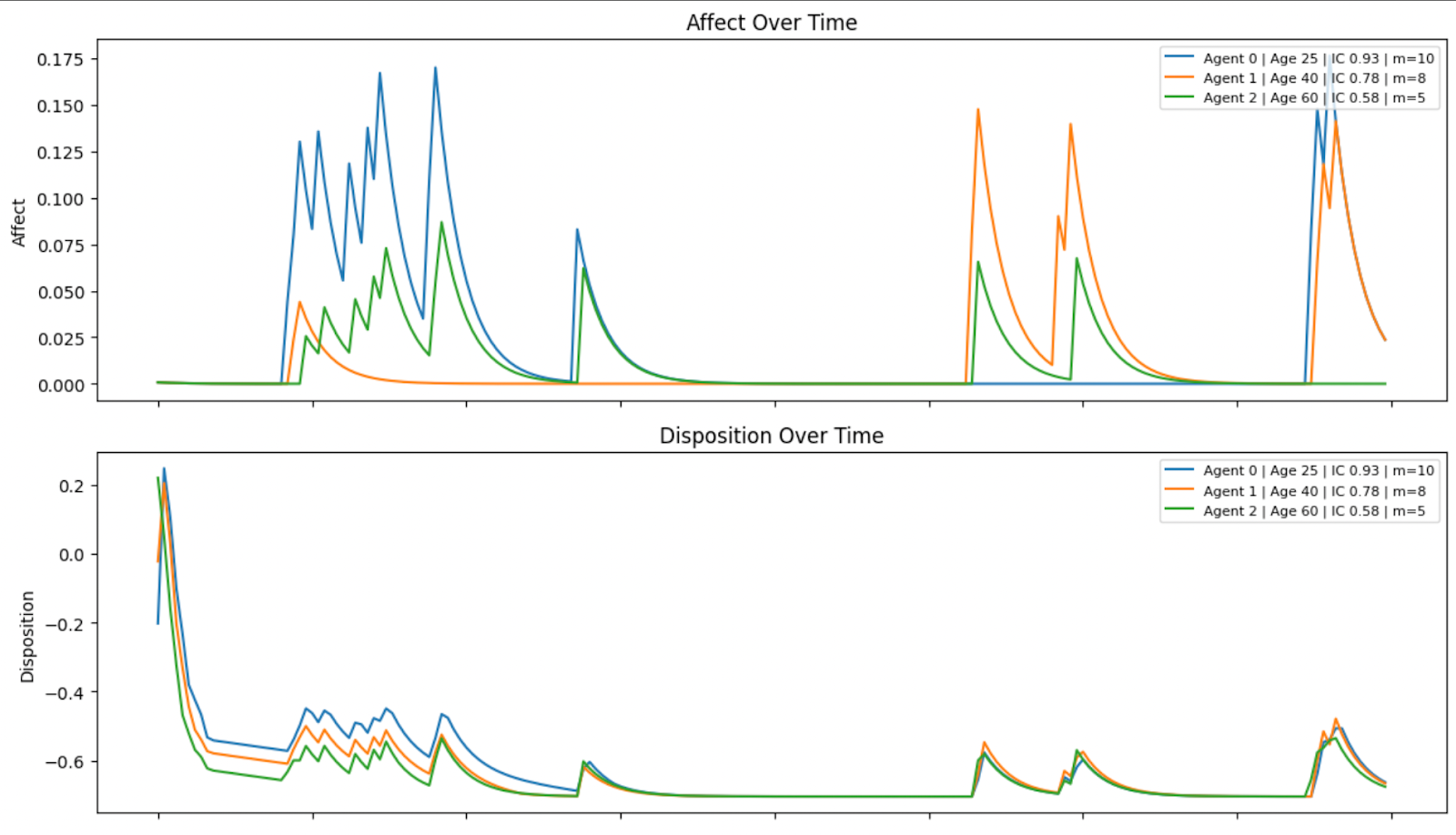}
    \caption{Disposition and Affect with Temporal Memory Over Time}
    \label{fig4_a}
  \end{subfigure}
  \hfill
  \begin{subfigure}[b]{0.7\textwidth}
    \includegraphics[width=\textwidth]{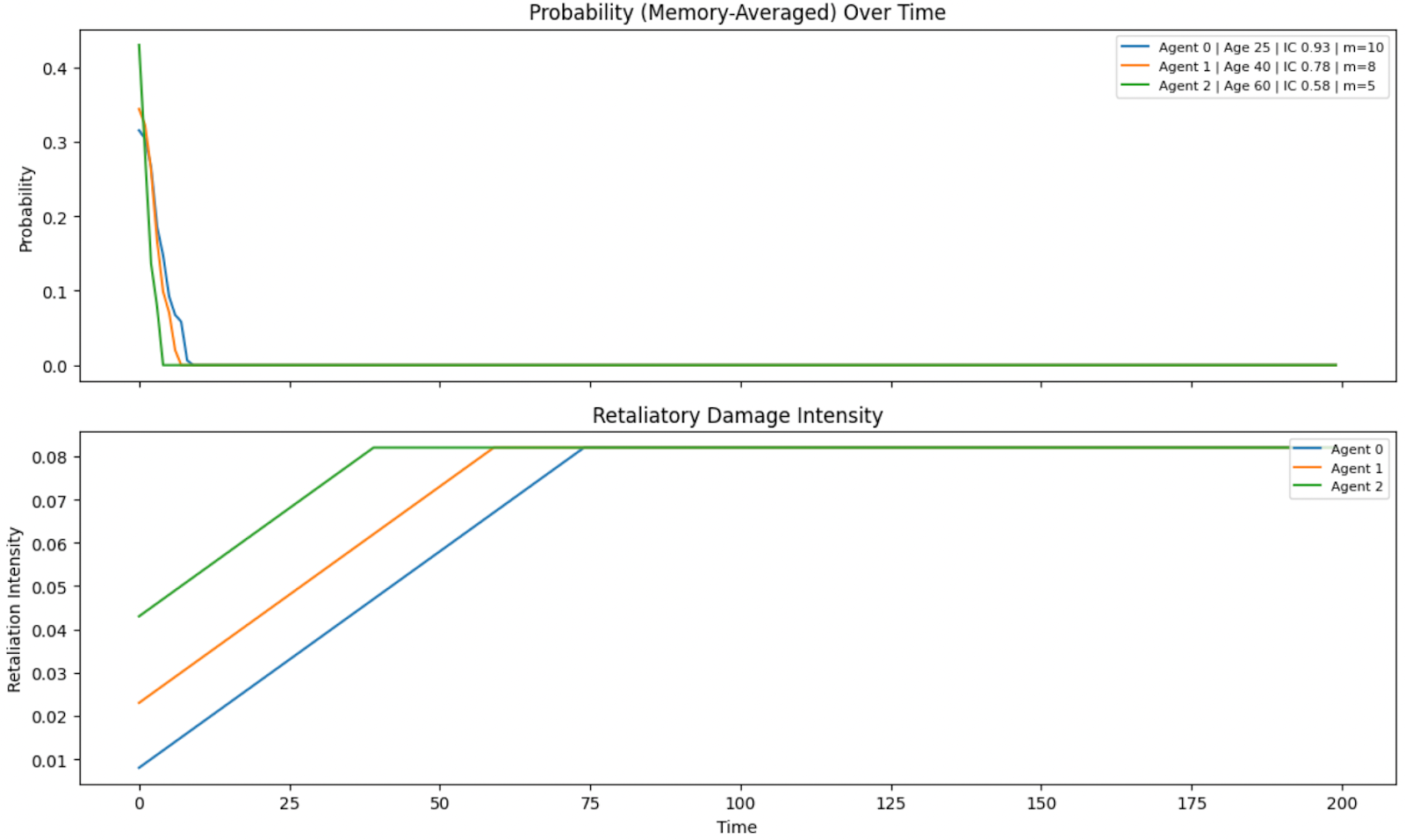}
    \caption{Memory-averaged Probability and Retaliatory Radius Over Time}
    \label{fig4_b}
  \end{subfigure}
  \caption{Temporal Memory Trends}
  \label{fig4}
\end{figure}

Introducing \textit{age-based impulse control} led to significant differences in activation patterns. Agents with lower impulse control (younger age) exhibited more frequent and earlier spikes in disposition, while older agents activated less often. This aligns with psychological expectations about youth-driven volatility. Figure~\ref{fig3} shows sharp peaks in disposition for younger agents and a flatter curve for older, more controlled individuals.

By allowing the \textit{radius of destruction} to scale with affect, we observed sudden, large-scale outbreaks of violence initiated by highly aroused agents. These agents impacted a wider spatial area during escalation events. Figure~\ref{fig3} also demonstrates how disposition peaks became sharper but shorter-lived, suggesting intense but episodic episodes of unrest.

\begin{figure}
\centering
\includegraphics[width=0.60\textwidth]{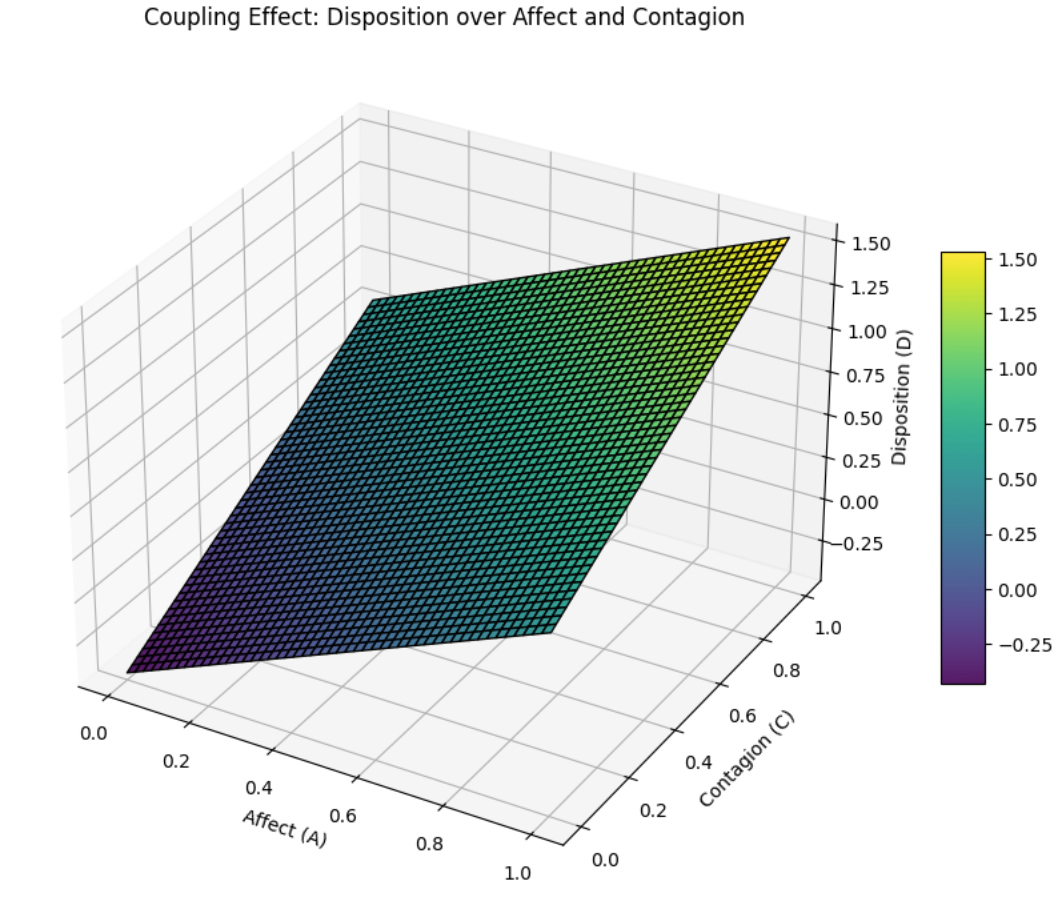}
\caption{Coupling Effect over Affect and Contagion.} \label{fig5}
\end{figure}

Adding a \textit{flight behavior} mode for agents with high perceived risk but low disposition significantly reduced spatial damage. These agents relocated instead of attacking, resulting in fewer destroyed patches and slower escalation. Figure~\ref{fig4_a} shows comparative disposition trajectories, revealing that flight behaviors preserved more of the environment, while fight-prone agents caused rapid spatial spread of destruction. Combined with Figure~\ref{fig2}, this illustrates the contrast between destructive and avoidance strategies.

\begin{figure}[htbp]
  \centering
  \begin{subfigure}[b]{0.7\textwidth}
    \includegraphics[width=\textwidth]{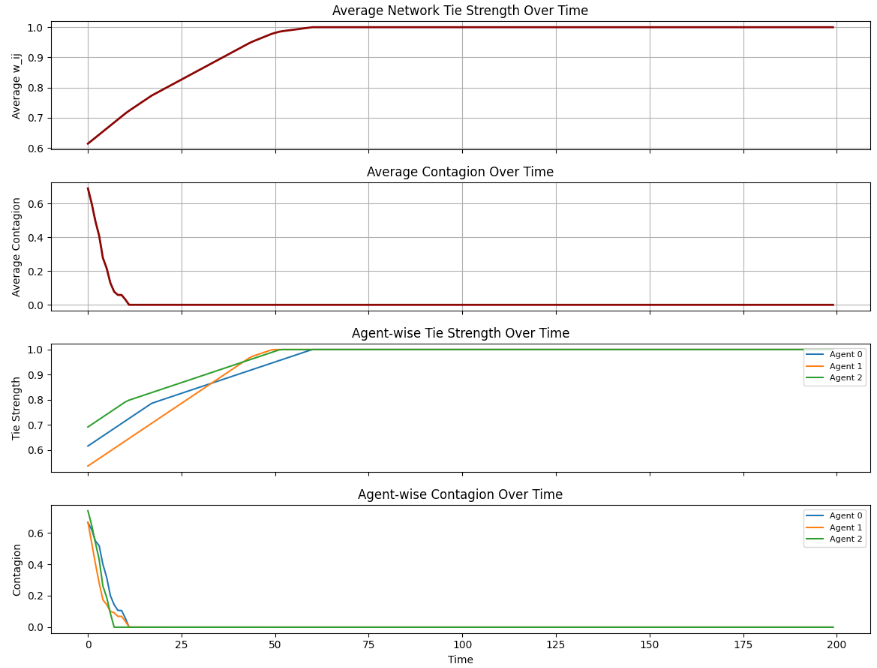}
    \caption{Without Uniform Shocks}
    \label{fig6_a}
  \end{subfigure}
  \hfill
  \begin{subfigure}[b]{0.7\textwidth}
    \includegraphics[width=\textwidth]{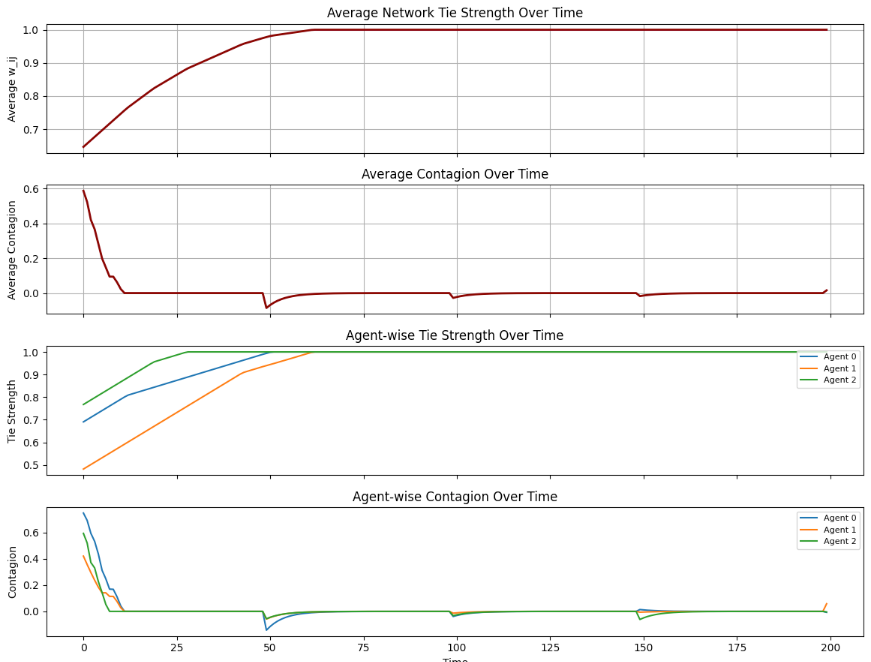}
    \caption{With Uniform Shocks}
    \label{fig6_b}
  \end{subfigure}
  \caption{Average Network Tie Strength and Contagion over time
}
  \label{fig6}
\end{figure}

We tested memory lengths $m = 3$ and $m = 12$ to examine how sustained perception of risk affects behavior. Agents with longer memory retained emotional responses well beyond the disappearance of local threats, while those with short memory returned to baseline affect levels more quickly. Figure~\ref{fig4_a} shows that Agent 2 (with long memory) sustains high disposition, while Agent 0 (with short memory) returns to low affect. Figure~\ref{fig4_b} shows how probability estimates evolve with memory size, confirming that longer memory increases persistence and escalation likelihood.

Disposition peaks occurred only when both \textit{affect and contagion} were simultaneously elevated, demonstrating a tipping-point effect. This is visualized in the diagonal activation band in Figure~\ref{fig5}, where disposition values sharply increase when both variables cross a threshold. The result supports the idea that emotional arousal alone is insufficient—social reinforcement is necessary to trigger action.

We tracked how tie strengths evolved under conditions with and without uniform shocks. In calm environments, emotionally similar agents converged early and formed \textit{stable affective clusters}. When periodic shocks were introduced, tie strength temporarily declined but eventually re-synchronized. Figure~\ref{fig6_a} shows convergence in stable settings, while Figure~\ref{fig6_b} captures divergence and recovery under external perturbation.

\begin{figure}[htbp]
  \centering
  \begin{subfigure}[b]{0.7\textwidth}
    \includegraphics[width=\textwidth]{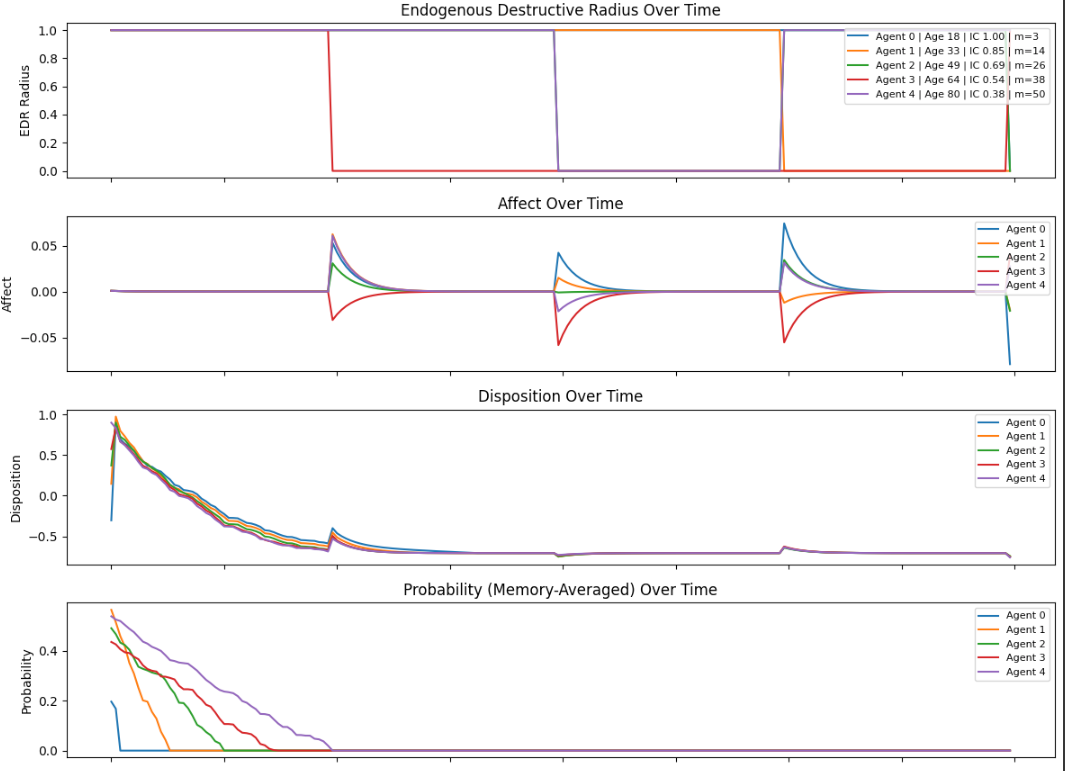}
    \caption{}
    \label{fig7_a}
  \end{subfigure}
  \hfill
  \begin{subfigure}[b]{0.7\textwidth}
    \includegraphics[width=\textwidth]{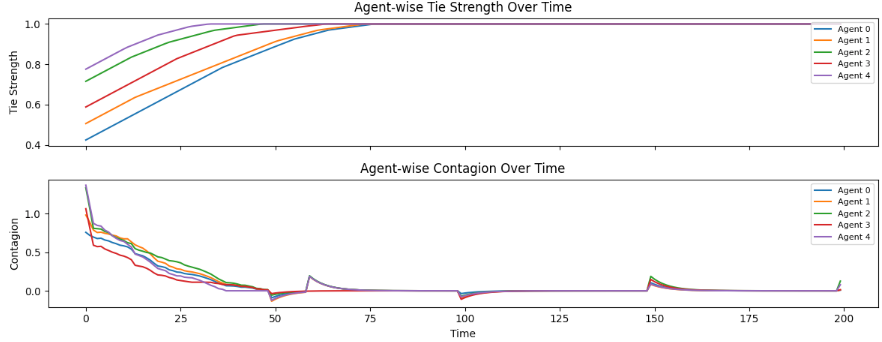}
    \caption{}
    \label{fig7_b}
  \end{subfigure}
  \caption{Overall Trends with Multiple Agents}
  \label{fig7}
\end{figure}

In a five-agent simulation, the interaction between affect, memory, and impulse control generated \textit{synchronized disposition spikes} and extended escalation periods. Retaliatory damage and contagion dampened over time, with affective synchrony driving early escalation. Figures~\ref{fig7_a} and \ref{fig7_b} visualize how system-wide coordination and individual heterogeneity jointly shape the duration and spatial extent of collective unrest.

\section{Conclusions}
This study advances computational modeling of collective behavior by integrating emotional dynamics, cognitive heuristics, and social learning into a unified agent-based framework. By extending the Agent\_Zero model with affective and probabilistic components, we demonstrate how slight variations in internal states—such as age-based impulse control, memory length, or affect-contagion coupling—can lead to significant divergences in macro-level outcomes. Our simulations reveal that crowd behavior is not merely the sum of environmental stimuli or rational incentives, but is profoundly shaped by internal affective states and the structure of social influence.

\begin{table}[ht]
\centering
\resizebox{0.95\textwidth}{!}{%
\begin{tabular}{|p{3.5cm}|p{3.2cm}|p{5.2cm}|p{2.5cm}|}
\hline
\textbf{Simulation Condition} & \textbf{Cognitive Basis} & \textbf{Key Observation} & \textbf{Figure Reference} \\
\hline
Baseline (3 agents) & Contagion Theory \cite{LeBon1895} & Localized destruction with short affect spikes; passive agents stay inert & Fig.~\ref{fig2} (Basic model) \\
\hline
Impulse Control (age variation) & Developmental Psychology \cite{Mischel2011} & Younger agents activate earlier and more frequently; higher affect volatility & Fig.~\ref{fig3}a \\
\hline
Endogenous Destructive Radius & Emotional Arousal and Behavioral Reactivity \cite{ledoux1996emotional} & High-affect agents cause wider spatial damage; sharper, shorter disposition peaks & Fig.~\ref{fig3}b \\
\hline
Fight vs. Flight Mode & Fear Response \cite{deGelder2004} & Flight reduces destruction, increases mobility; fight escalates damage & Fig.~\ref{fig2}, Fig.~\ref{fig4_a} \\
\hline
Retaliatory Damage & Feedback Loops \cite{Zillmann1975} & Introduces self-regulation; peer reactivity amplifies affect volatility & --- \\
\hline
Temporal Memory (long vs. short) & Episodic Learning \cite{bouton2004context} & Longer memory sustains emotional activation; short memory decays quickly & Fig.~\ref{fig4_a}, Fig.~\ref{fig4_b} \\
\hline
Coupling of Affect and Contagion & Dual Process Theory \cite{cacioppo2007social} & Disposition spikes only when both affect and contagion are high (tipping point) & Fig.~\ref{fig5} \\
\hline
Affective Homophily (with/without shocks) & Social Network Theory \cite{barabasi1999emergence} & Stable emotional clusters form; external shocks cause temporary destabilization & Fig.~\ref{fig6_a}, Fig.~\ref{fig6_b} \\
\hline
5-Agent Model Expansion & System Scaling \cite{baryam2002complexity} & Synchrony in affect; escalation cascades; early contagion, later dampening & Fig.~\ref{fig7} \\
\hline
\end{tabular} 
}
\vspace{0.5em}
\caption{Summary of Observed Effects from Simulation Results}
\label{tab:summary}
\end{table}

A key takeaway from this work is that the way agents respond to threats, whether they choose to fight or flee, can create very different patterns of unrest. Adding a "flight" option alongside "fight" revealed that even simple changes in response strategies can shape whether a society sees scattered avoidance or widespread destruction. Likewise, agents with longer memories tend to hold onto emotional reactions longer, making prolonged conflict more likely. These findings highlight the importance of including time and psychological nuance in models that aim to capture protests, riots, or other social unrest. We summarize these results in Table ~\ref{tab:summary}.

We also found that feedback loops, such as retaliatory damage and changing social ties, play a major role. When agents with similar emotional states cluster together, the system tends to stabilize unless a shock throws things out of balance, in which case new waves of unrest can flare up. This stop-and-start behavior reflects what we often see in real-world protests: stretches of calm followed by sudden outbreaks.

Looking at the stability of the system, we found that combinations of high impulse control, short memory, and loose social ties tend to keep things calm. On the flip side, low impulse control, strong emotional alignment, and long memory can lead to prolonged conflicts and social breakdowns. These insights could help policymakers test how resilient certain institutions or social structures are to pressure.

More broadly, this study argues that to truly understand collective behavior, we must consider what's happening inside the minds of individuals: their emotions, memories, and connections to others. Emotions and cognition are not peripheral—they are central to the decisions people make about whether or not to engage in collective action. The enhanced Agent\_Zero framework provides a useful platform for testing theories of political violence, emotional contagion, and the psychological underpinnings of group behavior.

Although the current model emphasizes individual-level cognitive variation, future work should address social science questions more directly. One direction is to incorporate narrative framing, allowing agents to respond not just to their environment but also to the stories and messages they encounter. This would allow for studying how misinformation, rumors, or propaganda spread and shape collective outcomes. Another promising extension involves modeling interactions with institutional agents, such as media or law enforcement, to examine how suppression of information or biased enforcement might influence unrest.

We could also introduce agent roles such as protester, bystander, or authority figure, enabling dynamic role shifts based on context. This would enable the exploration of dynamics such as discrimination, group identity, and social polarization. Such extensions would greatly improve the relevance of the model for real-world applications, be it in detecting early signs of unrest, evaluating institutional resilience under emotional stress, or comparing different communication strategies. By weaving in these social dimensions, AgentZero++ could evolve into a more robust platform for both academic inquiry and policy simulation of emotionally charged collective behavior.

\section{Acknowledgements}

We thank the reviewers and organizers of the CSSSA conference for their thoughtful feedback and support. This research was enriched by conversations with colleagues in the computational social science community and benefited from open-source modeling tools, particularly the Mesa framework. We are especially grateful to Joshua M. Epstein, whose Agent\_Zero framework served as the intellectual foundation for this project and continues to inspire new approaches in generative social science. We also gratefully acknowledge Dr. Robert Axtell for his invaluable guidance and intellectual support throughout the development of this work.

Some sections of this manuscript were drafted or refined using AI-assisted writing tools, including OpenAI’s ChatGPT and Grammarly. These tools were used to enhance clarity, organization, and readability, and all content was critically reviewed and edited by the author. All remaining errors are our own.

\bibliographystyle{splncs04}
\bibliography{mybibliography}

\begin{thebibliography}{10}
\providecommand{\url}[1]{\texttt{#1}}
\providecommand{\urlprefix}{URL }
\providecommand{\doi}[1]{https://doi.org/#1}

\bibitem{Bandura1999}
Bandura, A.: Moral disengagement in the perpetration of inhumanities.
  Personality and Social Psychology Review  \textbf{3}(3),  193--209 (1999)

\bibitem{baryam2002complexity}
Bar-Yam, Y.: Complexity rising: From human beings to human civilization, a
  complexity profile. New England Complex Systems Institute (2002),
  \url{https://necsi.edu/complexity-rising}

\bibitem{barabasi1999emergence}
Barab{\'a}si, A.L., Albert, R.: Emergence of scaling in random networks.
  science  \textbf{286}(5439),  509--512 (1999)

\bibitem{Baranski2010}
Baranski, J.V., Petrusic, W.M.: Aggregating conclusive and inconclusive
  information: data and a model of evaluative integration. Journal of
  Behavioral Decision Making  \textbf{23}(4),  383--403 (2010).
  \doi{10.1002/bdm.663}

\bibitem{Bloom2006}
Bloom, F.E., Lazerson, A., Nelson, C.A.: Brain, Mind and Behavior. Macmillan
  Learning, New York, 3 edn. (2006)

\bibitem{bouton2004context}
Bouton, M.E.: Context and behavioral processes in extinction. Learning \&
  memory  \textbf{11}(5),  485--494 (2004)

\bibitem{Browning1998}
Browning, C.R.: Ordinary Men: Reserve Police Battalion 101 and the Final
  Solution in Poland. HarperCollins, New York (1998)

\bibitem{cacioppo2007social}
Cacioppo, J.T., Amaral, D.G., Blanchard, J.J., Cameron, J.L., Carter, C.S.,
  Crews, D., Fiske, S., Heatherton, T., Johnson, M.K., Kozak, M.J., et~al.:
  Social neuroscience: Progress and implications for mental health.
  Perspectives on Psychological Science  \textbf{2}(2),  99--123 (2007)

\bibitem{cacioppo1982need}
Cacioppo, J.T., Petty, R.E.: The need for cognition. Journal of personality and
  social psychology  \textbf{42}(1), ~116 (1982)

\bibitem{Canetti1984}
Canetti, E.: Crowds and Power. Straus and Giroux, New York (1984)

\bibitem{cunningham2012motivational}
Cunningham, W.A., Brosch, T.: Motivational salience: Amygdala tuning from
  traits, needs, values, and goals. Current Directions in Psychological Science
   \textbf{21}(1),  54--59 (2012)

\bibitem{Epstein1997}
Epstein, J.M.: Nonlinear Dynamics, Mathematical Biology, And Social Science:
  Wise Use Of Alternative Therapies. Santa Fe Institute Studies in the Sciences
  of Complexity, Perseus / CRC Press, Reading, MA, 1 edn. (1997)

\bibitem{Epstein1999}
Epstein, J.M.: Agent‑based computational models and generative social
  science. Complexity  \textbf{4}(5),  41--60 (1999).
  \doi{10.1002/(SICI)1099-0526(199905/06)4:5<41::AID-CPLX9>3.0.CO;2-F}

\bibitem{Epstein2002}
Epstein, J.M.: Modeling civil violence: An agent-based computational approach.
  Proceedings of the National Academy of Sciences  \textbf{99}(Suppl. 3),
  7243--7250 (May 2002). \doi{10.1073/pnas.092080199}, epub May 7, 2002

\bibitem{Epstein2013}
Epstein, J.M.: Agent\_Zero: Toward Neurocognitive Foundations for Generative
  Social Science. Princeton University Press, Princeton, NJ (2013)

\bibitem{flache2017models}
Flache, A., M{\"a}s, M., Feliciani, T., Chattoe-Brown, E., Deffuant, G., Huet,
  S., Lorenz, J.: Models of social influence: Towards the next frontiers.
  Jasss-The journal of artificial societies and social simulation
  \textbf{20}(4), ~2 (2017)

\bibitem{deGelder2004}
de~Gelder, B., Snyder, J., Greve, D., Gerard, G., Hadjikhani, N.: Fear fosters
  flight: a mechanism for fear contagion when perceiving emotion expressed by a
  whole body. Proceedings of the National Academy of Sciences
  \textbf{101}(47),  16701--16706 (2004). \doi{10.1073/pnas.0407042101}, epub
  November 16, 2004

\bibitem{LeBon1895}
Le~Bon, G.: The Crowd: A Study of the Popular Mind. Dover Publications,
  Mineola, NY (1895)

\bibitem{ledoux1996emotional}
LeDoux, J.: The Emotional Brain: The Mysterious Underpinnings of Emotional
  Life. Simon and Schuster (1996)

\bibitem{Milgram1974}
Milgram, S.: Obedience to Authority: An Experimental View. Harper \& Row, New
  York (1974)

\bibitem{Mischel2011}
Mischel, W., Ayduk, O., Berman, M., Casey, B., Gotlib, I., Jonides, J., Kross,
  E., Teslovich, T., Wilson, N., Zayas, V., Shoda, Y.: “willpower” over the
  life span: Decomposing self-regulation. Social Cognitive and Affective
  Neuroscience  \textbf{6}(2),  252--256 (2011)

\bibitem{Nietzsche2002}
Nietzsche, F.: Beyond Good and Evil: Prelude to a Philosophy of the Future.
  Cambridge University Press (2002), original work published 1886

\bibitem{Zillmann1975}
Zillmann, D., Bryant, J., Cantor, J.R., Day, K.D.: Irrelevance of mitigating
  circumstances in retaliatory behavior at high levels of excitation. Journal
  of Research in Personality  \textbf{9}(4),  282--293 (1975).
  \doi{10.1016/0092-6566(75)90003-3}

\bibitem{Zimbardo2007}
Zimbardo, P.: The Lucifer Effect: Understanding How Good People Turn Evil.
  Random House, New York (2007)

\end{thebibliography}

\end{document}